\documentclass[12pt]{iopart}
\usepackage{iopams}
\usepackage{graphicx}
\usepackage{amssymb}
\newcommand{\bra}[1]{\, \langle\,{#1} \,|\,}
\newcommand{\ket}[1]{\,| \, {#1} \,\rangle  \,}
\newcommand{\braket}[2]{\, \langle\,{#1} \, | \, {#2} \,\rangle  \,}
\newcommand{\MyRe}{{\cal R} e \,}  
\newcommand{\MyIm}{{\cal I} m \,}    

\begin{document}
\title{Broadband noise decoherence in solid-state complex architectures}

\author{E Paladino$^1$,  A D'Arrigo$^1$, A Mastellone$^{2,1}$  and G Falci$^1$}

\address{$^1$  Dipartimento di Metodologie Fisiche e 
Chimiche (DMFCI), Universit\'a di Catania. Viale A. Doria 6, 
95125 Catania  (Italy) \& MATIS CNR - INFM, Catania .\\
$^2$ C.I.R.A. Centro Italiano Ricerche Aerospaziali -
Via Maiorise snc - 81043 Capua, CE (Italy)}

\ead{epaladino@dmfci.unict.it}

\begin{abstract}
Broadband noise represents a severe limitation towards the 
implementation of a solid-state quantum information processor.
Considering common spectral forms, 
we propose a classification of noise sources based on 
the effects produced instead of on their microscopic origin.
We illustrate a multi-stage approach to 
broadband noise 
which systematically 
includes only the relevant information on the environment, 
out of the huge parametrization needed for a microscopic description.
We apply this technique to a solid-state two-qubit gate in
a fixed coupling implementation scheme.
\end{abstract}

\submitto{\PS}

\section{Introduction}

Building scalable multi-qubit systems is presently the main challenge
towards the implementation of a solid-state quantum information processor~\cite{Nielsen}.
High-fidelity single qubit gates based both on semi-~\cite{single-semic} 
and super-conducting technologies are nowadays available. In particular, 
for the three basic types of superconducting qubits (charge, flux and phase) 
single-qubit operations with high quality factor have been demonstrated in different 
laboratories~\cite{single-super,kn:vion}. Further improvement has been
recently achieved with circuit-QED architectures~\cite{trasmon,circuitQED}.   
Multi-qubit systems instead have been proved harder to operate, the
main limitation arising from the broadband solid state noise. 
The requirements for building an elementary quantum processor are 
in fact quite demanding on the efficiency of the protocols.
This includes both severe constraints on readout
and a sufficient isolation from fluctuations
to reduce decoherence effects.

Based on experiments on the different Josephson junction (JJ) setups, 
there is presently a general consensus on the most common noise spectral forms 
and on the main consequences on systems evolutions.
Typically noise is broadband and structured, i. e. the noise spectrum extends 
to several decades, it is non-monotonic, sometimes a few resonances are present. 

Noise with  $1/f$ spectrum is  common to virtually all nanodevices.
Its physical
origin varies from device to device and depends on the specific material. Different 
implementations are in fact
more sensitive either to charge, flux or to critical current fluctuations 
with spectral density scaling as the inverse of the frequency.  
The presence of slow components in the environment makes 
the decay of the coherent signal strongly dependent on 
the experimental protocol being
used~\cite{kn:vion,kn:nakamura-echo,kn:falci-varenna,PRL05}. 
Measurements protocols requiring numerous repetitions are particularly
sensitive to the unstable device calibration due to low-frequency
fluctuations. The leading effect is defocusing of the measured signal,  analogous
to inhomogeneous broadening in NMR~\cite{slitcher}.
The intrinsic high-frequency cut-off of  $1/f$ noise is hardly detectable, measurements
typically extending to $100$~Hz (recently charge noise up to $10$~MHz has been 
detected in a SET~\cite{delsing08}).  
Incoherent energy exchanges between system and environment, leading to relaxation and
decoherence, occur at typical operating frequencies (about $10$ GHz).
Indirect measurements of noise spectrum in this frequency range often 
suggest a white or ohmic behavior~\cite{nak-spectrum,ithier}.
In addition,  narrow resonances at selected frequencies (sometimes 
resonant with the nanodevice relevant energy scales) have being
observed~\cite{martinis04,mooji}.  
In certain devices they originate from the circuitry~\cite{Delft} and may eventually be
reduced by circuit design. More often, resonances are signatures of the
presence of spurious fluctuators which also show up in the
time resolved evolution, unambiguously proving the discrete nature of 
the noise sources~\cite{kn:duty}. Such fluctuators may severely limit the reliability
of nanodevices~\cite{PRL05,Altshuler}.

Explanation of this rich physics is beyond  phenomenological theories 
describing the environment as a  set of harmonic 
oscillators.
On the other side, an accurate characterization 
of the noise sources might be a priory inefficient, since
a huge number of parameters would be required for a microscopic description. 
Therefore one may follow a different route, consisting in
classifying noise sources on the basis of their effects instead of on their nature
and to understand, case by case, which is the efficient description of the environment.
The required information may in fact depend on the specific protocol,
especially when the environment is long-time correlated.  
This program is meaningful for quantum information where
relevant time scales are much smaller than the decoherence time. This means that
in the generic favorable situation coupling with the environment has simple effects
on the system dynamics. 
As a difference with condensed matter physics where long time behavior is 
emphasized and interesting problems involve entanglement of a many-body system.

Here we illustrate  a road-map to treat broadband noise which allows 
to obtain reasonable approximations by systematically 
including only the relevant information on the environment, 
out of the huge parametrization needed to specify it.
The  multi-stage approach for the different classes of broadband noise has 
been originally introduced for single qubit gates~\cite{PRL05}. 
The obtained predictions for the  decay of the coherent signal are 
in agreement with observations in various JJ implementations and in  
different protocols~\cite{kn:falci-varenna,PRL05}. We mention the
observed decay of Ramsey fringes in charge-phase qubits~\cite{kn:vion}
and recent results on flux/phase qubits~\cite{poletto}.
Here we extend the procedure to complex solid-state architectures. 
As an illustrative case, we perform a systematic analysis
of the effects and interplay of low and high frequency noise components
in a two-qubit gate in a fixed coupling scheme.
Such a systematic analysis 
points out that efficient operations in the solid state require
an accurate preliminary characterization of the noise spectral characteristics and
tuning appropriately the device working point.  

The paper is organized as follows: in Section 2 we introduce the general
framework of solid-state multi qubit gates and illustrate the characteristics of
the most common spectral forms. In Section 3 we propose a classification of the
noise sources and in Section 4 we present a multi-stage approach to deal
with broadband noise. Section 5 is dedicated to adiabatic noise. In Section 6
the multistage approach is applied to a universal two-qubit gate. Section 7 summarizes
our main findings.

\section{Multi-qubit systems and noise}

A multi-qubit gate can be modeled by the following Hamiltonian ($\hbar=1$)
\begin{equation}
{\mathcal H}_G(t) =  \sum_i \,  
{\mathcal H}_Q^{(i)}(t)
\, + \, 
\sum_{i,j} {\mathcal H}_{ij}(t)
\end{equation}
where for each qubit, labeled by  $i=1, \dots, n$, 
${\mathcal H}_Q^{(i)}(t) = -\frac{1}{2}  \; \vec \Omega_i(t) \cdot \vec \sigma^{(i)} $
includes set of parameters intrinsic to the device and time dependent classical 
control fields. 
Manipulation of tunable fields allows control of the dynamics and design of arbitrary
single qubit operations via unitary transformations. Interactions among qubits
and the needed additional parametrization to describe the multi-qubit gate are
included in  $\sum_{i,j} {\mathcal H}_{ij}(t)$.  Depending on the design, the
control Hamiltonian may span a reduced subspace of the qubit Liouville space. In
other words, there is a limited number of ports available for control, for tuning, 
for state preparation and measurement. Noise is also coupled via these ports.
It is rather usual that only one of the control fields allows the required 
fast addressing of each qubit. Therefore we focus on a model where both the control fields,
$A(t)$, and the environment are coupled to a single qubit operator, say $\sigma_z^{(i)}$,
\begin{eqnarray}
{\mathcal H}_Q^{(i)} (t)
\, &=& \,  -\frac{\Omega_i}{2} \, \cos \theta_i  \, \sigma_z^{(i)} \, 
-\frac{\Omega_i}{2} \,
\, \sin \theta_i \,  \sigma_x^{(i)}  -\frac{1}{2}\, A(t) \, \sigma_z^{(i)} \, ,
\label{Hamqubit} \\
{\mathcal H}(t)  &=& \,{\mathcal H}_G(t) + \frac{1}{2} \, \sum_i \sigma_z^{(i)} \otimes \hat X_i +
{\mathcal H}_R \, ,
\label{H}
\end{eqnarray}
where 
the polar angles $\theta_i$ define qubit-$i$ 
working point. 
The  environment  Hamiltonian is $ {\mathcal H}_R$ and $ \hat X_i$  is a collective environment
variable acting on qubit $i$. 
Model (\ref{H}) implies a projection of the device Hamiltonian onto the
subspace spanned by the two lowest energy eigenstates for each qubit. 
This description is valid provided that manipulations with the control fields do not
induce leakage to higher energy states of the device~\cite{siewert}.

We leave unspecified the nature of the noise sources described by ${\mathcal H}_R$,
relevant cases being either discrete or Gaussian fluctuations.
In addition, the environment Hamiltonian may
include correlations among noise sources affecting each qubit.  
In the spirit of the present analysis, we assume that the 
only information at our disposal is the power spectrum of $\hat X_i$ fluctuations 
\begin{equation}
S_{X_i}(\omega) = \int_0^\infty dt \; e^{i \omega t} \left\{ \frac{1}{2} \,
\langle \hat X_i(t) \hat X_i(0) + \hat X_i(0) \hat X_i(t) \rangle - \langle \hat X_i
\rangle^2 \right \}
\, ,
\end{equation}
where $\langle ... \rangle$ denotes the equilibrium average with respect
to $ {\mathcal H}_R$.
We remark that the effect of the environment on the system dynamics is completely 
characterized by $S_{X_i}(\omega) $
only if the environment
is composed of harmonic oscillators or if it is weakly coupled to the system and
short time correlated. Of course this is not the case if low-energy excitations determine
memory effects. This is the typical situation in the solid state where in general 
additional statistical information on the environment is required.

We assume that each noise component $\hat X_i$ has broadband spectrum 
$
S_{X_i}(\omega) = \frac{A}{\omega}$, 
$\omega \in \{\gamma_m, \gamma_M\} $
followed by a white or ohmic flank at frequencies $\omega \gg \gamma_M$.
Low- and high-frequency cutoffs 
depend on the specific setup. 
Impurities of various origin responsible for random telegraph fluctuations 
contribute Lorentzian peaks to the power spectrum 
$S_{rtn}(\omega)= \frac{v_0^2}{2} \, \frac{\gamma}{\gamma^2 + \omega^2}$.
Such a spectrum originates from classical fluctuations $\hat X \to X(t)$, where $X(t)$ 
randomly switches between two values $ \{0, v_0\}$ with rate $\gamma$. 
An ensemble of $N_{bi}$ bistable impurities with a distribution of switching rates
$\propto 1/ \gamma$, $\gamma\in [\gamma_m,\gamma_M]$ and average coupling
strength $\overline{v}$, gives rise to $1/f$-spectrum,
$S^{1/f}(\omega) \approx
 [(\pi/4) 
\,N_{bi} \overline{v^2} / 
\ln (\gamma_M/\gamma_m)]\,\,\omega^{-1}$~\cite{weissman}.  
Selected Lorentzian
peaks may be visible in the spectrum if individual impurities are strongly
coupled, i. e. when $v_0/\gamma \gg 1$.
Instead, damped coherent fluctuators at frequencies $\omega_{0}$, in the simplest cases contribute 
to the total spectrum with additional peaks, 
$S_{cf}(\omega)= \frac{v_0^2}{2} \, \frac{\gamma_0}{\gamma_0^2 + (\omega -
\omega_{0})^2}$. Spectroscopic evidence of coherent impurities of frequency $\omega_0$ close
to the qubit Larmor frequency has been reported in \cite{martinis04,coherent}.
Remarkably, the possibility to exploit spurious quantum two-level systems as
qubits~\cite{zagoskin} or for 
quantum memory operations has been recently demonstrated~\cite{neeley08}. 
In these cases a quantum description of the impurity is required~\cite{PRB08,coherent-th}.

\section{Three classes of noise}

The above description illuminates that in the solid state we have to deal
with broadband and structured noise. In other words, the noise spectrum extends to
several decades, it is non-monotonic, sometimes a few resonances are present. 
The various noise sources  responsible for the above phenomenology
have a qualitative different influence on the system evolution. 
This naturally leads to a classification of the noise sources
according to the effects produced rather than to their specific nature.

The effects of high-amplitude noise at low frequencies, like $1/f$ noise, vary from 
protocol to protocol. This feature is typical of non-Markovian baths. 
Quantum operations necessarily require repetitions of single detections, each 
leading to Boolean answer. Therefore, even "single shot" measurements result 
from numerous repetitions of single runs in an overall process which 
may last minutes. In the presence of low frequency fluctuations this leads to
unstable device calibration and random clock frequencies in the various repetitions.
As a result, a de-focused signal is observed, a phenomenon analogous to
inhomogeneous broadening in NMR. Re-focusing protocols, like echo or some dynamical
decoupling schemes, allow partial recovery of  the 
signal~\cite{kn:nakamura-echo,ithier}. Since the environment
is long-time correlated, statistical information beyond the power
spectrum may be required to describe its effects. This is the case for instance in
echo protocols~\cite{kn:falci-varenna,Altshuler}. 
Environments with long-time memory belong to the class of adiabatic noise, for 
which the Born-Oppenheimer approximation is applicable.  We classify this part of the
noise spectrum as "adiabatic noise".

Noise at higher frequency, around the system typical scales (e.g. single qubit 
Larmor frequencies), results in incoherent energy exchanges between the quantum
device and the environment. In particular, high-frequency noise  is responsible for spontaneous decay. 
For systems relevant for quantum information, the system-bath
coupling is rather small, in addition high-frequency noise is short-time correlated.
Therefore, in simplest cases, effects can be described by a Born-Markov master
equation~\cite{kn:cohen}.
For single qubit gates it leads to the relaxation and secular dephasing times,
 $T_1= \sin^2 \theta \, S_{X}(\Omega)/2$ and $T_2 = 2 T_1$~\cite{slitcher}.   
 We classify this part of the noise spectrum as "quantum noise".

Finally, resonances in the spectrum unveil the presence of discrete noise sources
which severely effect the system performances, in particular reliability of devices.
This is the case when classical impurities are slow enough to induce a visible
bistable instability in the system intrinsic frequency. For instance,
single qubit gates in the presence of random telegraph noise (switching rate $\gamma$) 
may display two effective frequencies, $\Omega$ and $\Omega^\prime$, depending on the 
impurity state.
 Their visibility is measured by the ratio
$g=(\Omega^\prime - \Omega)/\gamma$. Beatings can be observed in the "strong coupling"
regime $g > 1$~\cite{PRL02}. Quantum impurities may also entangle with the device. This additionally
leads to a variety of  features, like peculiar temperature dependencies of decay rates~\cite{PRB08}.   
Under these conditions, knowledge of the power spectrum is absolutely insufficient and
in order to describe these effects the relevant system Hilbert space has to be
enlarged to include the responsible environmental degrees of freedom. 
Effects in general depend on the specific protocol and require a microscopic 
model of the fluctuators.   We classify this part of the
noise spectrum as "strongly-coupled noise".

Each noise class requires a specific approximation scheme, which is not
appropriate for the other classes. The overall effect results from the interplay
of the three classes of noise. In the following Section we will illustrate a 
multi-scale theory to deal with solid state broadband noise.

\section{Multi-scale theory for broadband noise}

We are interested to a reduced description of the $n$-qubit system, 
expressed by the reduced density matrix $\rho^{n}(t)$. It is formally obtained
by tracing out environmental degrees of freedom from the total density matrix 
$W^{Q,A,SC}(t)$, which depends on  quantum (Q), adiabatic (A) and strongly coupled 
(SC) variables.
The elimination procedure can be conveniently performed by separating
in the interaction Hamiltonian, $\sum_i \sigma_z^{(i)} \otimes \hat X_i$, various
noise classes, e.g. by formally writing
\begin{equation}
\sigma_z^{(i)} \otimes \hat X_i = \sigma_z^{(i)} \otimes \hat X_i^Q + 
\sigma_z^{(i)} \otimes \hat X_i^A + \sigma_z^{(i)} \otimes \hat X_i^{SC} \, .
\label{int-plit}
\end{equation}
Adiabatic noise is typically correlated on a time scale much longer than 
the inverse of the natural frequencies $\Omega_i$, 
then application of the Born-Oppenheimer approximation
is equivalent to replace $\hat X_i^A$ 
with a classical stochastic field  $X_i^A(t)$.
This approach is valid when the contribution of adiabatic noise to 
spontaneous decay is negligible, a necessary condition being 
$t \ll T_1^A \propto S_{X}^A(\Omega_i)^{-1}$. This condition is usually satisfied at
short enough times, 
since $S_{X}^A(\omega)$ is substantially  different from zero only at frequencies 
$\omega \ll \Omega_i$.
 
This fact already suggests a route to trace-out different noise classes
in the appropriate order.  The total density matrix parametrically depends on 
the specific realization of the slow random drives $\vec X(t) \equiv \{X_i^A(t)\}$ and 
may be
written as $W^{Q,A,SC}(t)= W^{Q,SC}(t | \vec X(t))$. 
The first step is to trace out quantum noise. In the simplest cases this requires
solving a master equation. In a second stage, the average over all the realizations of
the stochastic processes, $\vec X(t)$, is performed. This leads to a reduced density
matrix for the $n$-qubit system plus the strongly coupled degrees of freedom.
These have to be traced out in a final stage  by solving the Heisenberg
equations of motion, or by approaches suitable to the specific microscopic 
Hamiltonian or interaction. For instance, the dynamics may be solved exactly
for some special quantum impurity models at pure dephasing, $\theta_i=0$, 
when impurities are longitudinally coupled to each qubit~\cite{Altshuler,PRL02}.  
The ordered multi-stage elimination procedure can be formally written as
\begin{equation}
\rho^{n}(t) = Tr_{SC} \left \{ 
\int {\mathcal D}[\vec X(t)] \,  P[\vec X(t)] \; 
Tr_{Q} \Big[ \, W^{Q,SC}\Big(t | \vec X(t)\Big) \, \Big]\right\} \, .
\end{equation}
In the following Section we concentrate on the elimination of adiabatic noise. The
ordered procedure will be illustrated in Section 6 for a two-qubit gate.

\section{Adiabatic noise}

In general, the adiabatic approximation holds true for times short enough
to fulfill the necessary condition,  $t \ll T_1^A$. 
In the peculiar, pure dephasing regime, $\theta_i=0$, relaxation processes are 
forbidden  and the adiabatic approximation is exact for any $S_{X_i}(\omega)$.
In addition, the adiabatic scheme can be applied also in the presence of
correlations between processes $X_i(t)$ and $X_j(t)$~\cite{NJP-special}.  
For the sake of clarity here 
we consider adiabatic noise  affecting independently each qubit. Moreover,
we  exclude time-dependent drives in $A(t)$ and in $ {\mathcal H}_{ij}(t)$.
The procedure can be straightforwardly extended for instance to 
Rabi oscillations and other ac-drives~\cite{preparation}.

Suppose we are able to diagonalize, for each  realization $\vec X(t)$ of the 
stochastic processes, the system Hamiltonian, 
\begin{equation}
{\mathcal H}(t) = - \sum_i \,  \frac{1}{2}  \; \vec \Omega_i \cdot \vec \sigma^{(i)} \, + \, 
\sum_{i,j} {\mathcal H}_{ij} + \frac{1}{2} \, \sum_i \sigma_z^{(i)} \otimes  X_i(t) \, ,
\label{eq:hamiltonian-stochastic}
\end{equation}
and denote $\ket{m(\vec X_t)}$ 
an instantaneous eigenstate of (\ref{eq:hamiltonian-stochastic}) with eigenvalue
$E_m(t)$.  If the system is prepared in a pure state,
$\ket{\psi \,0}$, in the adiabatic approximation, each component 
$\ket{m(\vec X_0)}\!\!\braket{m(\vec X_0)}{\psi \,0}$  
evolves in time according to 
$$
\ket{m(\vec X_t)} {\mathrm{e}}^{i\Phi^D_m(t)}  \braket{m(\vec X_0)}{\psi \,0} \, ,
$$
where
 $\Phi^D_m(t) = - i \int_0^t ds \, E_m(s)$ 
is the dynamic phase. 
The stray fields $\vec X(t)$  affect the time evolution of the  gate fidelity 
via modifications of the phases $\Phi^D_m(t)$ and of the eigenstates $ \ket{m(\vec X_t)}$.
In addition, $ \ket{m(\vec X_0)}$ reflects imperfect preparation due to the random field. 
This occurs, for instance, when the system is prepared by a  
reset to a pure state vector $\ket{\psi \,0}$ followed by 
an initialization pulse whose effect depends also on the stray field at $t=0$. 
The evolution of the system conditional 
density matrix can be presented in a compact form
\begin{eqnarray}
\rho^n\big(t|\vec X(s)\big) 
&=&  
\sum_{m \, p } R_{m p}[\vec X_0, \vec X_t]
\;
{\mathrm e}^{- i \int_0^t d s \,
\Omega_{m p}(s)}  \, ,
\end{eqnarray}
where  the instantaneous splittings appearing in the phase factor are
$
\Omega_{mp}(t) =  -\, \Omega_{pm}(t) = E_m(t)- E_p(t)
$
and we have introduced the operator
\begin{equation} 
R_{m p}[\vec X_0, \vec X_t] =
\, \ket{m(\vec X_t)}\bra{m(\vec X_0)} 
\rho(0) \ket{p(\vec X_0)}
\bra{p(\vec X_t)}
\end{equation}
which contains information about the preparation and the eigenstates errors.
Finally, the average of $\rho^n\big(t|\vec X(s)\big)$ 
over the stochastic process yields the system density matrix, which can be presented as a path integral
\begin{eqnarray}
\label{eq:pathint-adiabatic}
\rho^n(t) 
&=&
\int \hskip-2pt {\cal D} \big[\vec X(s) \big] \,
P[\vec X(s)] \, \rho^n\big(t|\vec X(s)\big)  \, .
\end{eqnarray}
Here $P[\vec X(s)]$ contains information both on the stochastic 
processes and on details of the specific protocol. It is convenient 
to split it as follows
$$
P[\vec X(s)] \,=\, F[\vec X(s)] \; p[\vec X(s)] \, ,
$$
where $p[\vec X(s)]$ is the probability of the realization $\vec X(s)$ . 
The filter  function $F[\vec X(s)]$ describes the specific operation. For 
most of present day experiments on solid-state qubits 
$F[\vec X(s)]=1$. For an open-loop feedback protocol, which allows initial 
control of some collective variable of the environment,
say  $\vec X_0 = 0$, we should put $F[\vec X(s)] \propto \, \delta(\vec X_0) = \, \Pi_i \delta(X_{i0})$.

A critical issue is the identification of $p[\vec X(s)]$ 
for the specific  noise sources, as those displaying $1/f$ power spectrum.
If we sample the stochastic process at times $t_k = k \Delta t$, with   
$\Delta t = t/m$ and $k=0, \dots, m$, we can  identify
\begin{equation}
p[\vec X(s)] = \lim_{m \to \infty} 
p_{m+1}(\vec X_m, t;\dots; \vec X_1, t_1; \vec X_0, 0)  \, , 
\label{sampling}
\end{equation}
where $p_{m+1}(.)$ is a $m+1$ joint probability and we have used the
shorthand notation $\vec X_k \equiv \vec X_{t_k}$. 
In the following we will propose a systematic method to 
select only the relevant statistical information on the stochastic 
process out of the full characterization included in $p[\vec X(s)]$.

We would like to remark, at this point, that in the adiabatic treatment 
the word decoherence is perhaps abused. 
Decoherence is ultimately due to entanglement of the qubit 
to a quantum environment~\cite{kn:zurek,kn:palma96}, 
whereas classical adiabatic noise produces only de-focusing.
However, unless the signal can be {\em totally} re-focused, and 
this is impossible in practice, the behavior of $\rho(t)$ 
is the same as for true 
decoherence~\cite{kn:cory-et-al-1998}. In other words, 
although proper and 
improper mixed states at the fundamental 
level are well distinct concepts~\cite{kn:isham-95}, 
they are not distinct in the density matrix description.
We also observe that present day ``applied'' 
research on solid-state coherent nanodevices for quantum information focuses on 
the "short-time" dynamics, since a signal which is almost 
decayed is useless. In this context, methods as the adiabatic
approximation are valuable even if they are not accurate at 
long time scales and are not valid down to zero temperature.  

\subsection{Longitudinal approximation}

The longitudinal approximation consists in neglecting modifications of the 
eigenstates $ \ket{m(\vec X_t)}$ and preparation effects. 
Without loss of generality, we may assume vanishing average
of the stochastic processes after the  preparation pulse, $\vec X_0$.
The longitudinal approximation amounts to put
 $\ket{m(\vec X_t)} =  \ket{m(\vec X_0)} =  \ket{m}$ 
where $\ket{m}$ is an eigenstate of the system Hamiltonian
${\cal H}_G$. 
Then $R_{m p}[\vec X_0, \vec X_t] \approx 
\ket{m} \rho^n_{mp}(0) \bra{p}$ is a projected element of the initial density 
matrix and Eq.(\ref{eq:pathint-adiabatic}) simplifies to
\begin{eqnarray}
\label{eq:pathint-adiabatic-longitudinal}
\rho^n_{mp}(t) 
&=& \rho^n_{mp}(0) \, 
\int \hskip-2pt {\cal D}[\vec X(s)] \,
P[\vec X(s)] \; 
{\mathrm e}^{- i \int_0^t \!\!d s \, 
\Omega_{m p}(s)}  \, .
\end{eqnarray}
The significance of the longitudinal approximation is easily illustrated for
a single qubit gate, when the system Hamiltonian reduces to 
\begin{equation}
{\mathcal H}(t) = - \frac{\Omega}{2} \cos \theta \, \sigma_z \, - \frac{\Omega}{2} 
\, \sin \theta \,  \sigma_x - \frac{1}{2} \sigma_z \otimes X(t) \, .
\label{eq:hamiltonian-stochastic2}
\end{equation}
In this case, 
retaining  variations of the splitting  $\Omega$ amounts to consider
only  fluctuations of the length of the vector $\vec{\Omega}$. This would be the only
effect if the noise would act  longitudinally, $\theta=0$. 
Variations of the eigenstates $ \ket{m(X_t)}$ originate instead from 
"transverse" variations of $\vec{\Omega}$ .

The longitudinal assumption has been performed  in Ref.\cite{kn:makhlin-04}  
to discuss the effect of Gaussian  adiabatic environments. 
The present approach  automatically provides constraints on its validity  and 
shows that whereas errors due to transverse fluctuations 
are weakly dependent on time, phase errors accumulate.  Therefore,
transverse noise in the adiabatic approximation has possibly 
some effect only at very short times, but the phase 
damping channel eventually prevails. 
These considerations strongly  depend on the amplitude of the noise. We checked 
analytically and with simulations that they hold true for 
realistic figures of noise as those measured in experiments 
by Zorin et al.~\cite{kn:zorin}. 

In the longitudinal approximation, diagonal elements of the reduced density matrix
in the eigenstate basis, $\ket m = \ket \pm$, do not decay.
Instead, the decay of the off-diagonal elements results from 
$\rho_{\!\raisebox{-2pt}{\tiny$+-$}}(t) = \rho_{\!\raisebox{-2pt}{\tiny$+-$}}(0) 
\exp[- i \,\Phi(t)]$, where the  complex average  phase is given by 
\begin{equation}
\label{eq:blur}
\Phi(t) = 
-  \Omega t + i \ln 
\int\!\! 
{\cal D}[ X(s)] \, P[X(s)] \, 
{\mathrm e}^{i \int_0^t \!\!ds  \, \Omega[X(s)]} \, .
\end{equation}
We notice that the longitudinal approximation may be exact for certain protocols, 
which therefore are not affected by adiabatic transverse noise. For 
instance, this is the case if the system is described by (\ref{eq:hamiltonian-stochastic2}) 
and it is prepared in a state $\rho(0) = {1 \over 2} (1 \pm \sigma_y)$.
If   we measure $\sigma_y$, 
since $\bra{m(X_t)} \sigma_y \ket{p(X_t)}$ do not depend on $X_t$, 
the result is not affected by transverse fluctuations and 
$\overline{\langle \sigma_y(t) \rangle} = \langle \sigma_y(0) \rangle 
\, \cos[ \MyRe\!\Phi(t) ] \, \exp\{\MyIm\! \Phi(t)\}$. 
This the case of the decaying oscillations pattern measured with Ramsey 
interference~\cite{kn:vion}, if 
the effect of imperfect $\pi/2$ pulses is negligible.

\subsubsection{Static path approximation}
A standard approximation of the path integrals 
(\ref{eq:pathint-adiabatic},\ref{eq:pathint-adiabatic-longitudinal}) consists 
in neglecting the time dependence in 
the path, $\vec X(s) = \vec X_0$ and taking $F[\vec X]=1$. 
In this Static Path Approximation 
(SPA) the problem reduces to ordinary integrations with
$p_1(\vec X_0,0) \equiv p(\vec X_0)$. 
In the single qubit case, for instance, Eq.(\ref{eq:blur}) 
gives average  phase shift
\begin{eqnarray}
\label{eq:blur-static}
\Phi(t) \,\approx\, 
-  \Omega t + i \ln 
\int \hskip-2pt dX_0 \,
p(X_0) \;
{\mathrm e}^{i \Omega(X_0) t} \, ,
\end{eqnarray}
where $\Omega(X_0) =\sqrt{[\Omega \sin \theta + X_0]^2 + (\Omega \cos \theta)^2}$.
Eq.(\ref{eq:blur-static}) 
describes the effect of a distribution of stray energy 
shifts $\Omega_{m p}(X) - \Omega_{m p}(0)$ and corresponds to the
rigid lattice breadth contribution to inhomogeneous 
broadening~\cite{slitcher}. In experiments with solid state devices this 
approximation describes the measurement procedure consisting in 
signal acquisition and averaging over a large number $N$ 
of repetitions of the protocol, for an overall time $t_m$ 
(which may also be minutes in actual experiments). 
Due to slow fluctuations of the solid state
environment calibration, the initial value, 
$\Omega \sin \theta+X_0$,  fluctuates during the repetitions
blurring the average signal, independently on 
the measurement being single-shot or not. 

The probability $p(X_0)$ describes the 
distribution of the random variable obtained by sampling 
the stochastic process $X(t)$ at the initial time of each repetition, i. e.
at times $t_k = k \, t_m/N$, $k=0, N-1$. 
If $X_0$ results from many independent random variables 
of a multimode environment, the central limit theorem applies and
$p(X_0)$ is a Gaussian distribution with standard 
deviation $\sigma$
$$
\sigma^2 \,=\, \langle X^2 \rangle \,=\, \int \frac {d \omega}{\pi} \,
S_X^A(\omega) \, ,
$$
with integration limits $1/t_{m}$, 
and the high-frequency cut-off of the
$1/f$ spectrum, $\gamma_M$.
In the SPA the distribution standard deviation,
$\sigma$, is the only adiabatic noise characteristic parameter.
If the equilibrium average of the stochastic process vanishes, 
Eq.(\ref{eq:blur-static}) 
reduces to 
\begin{equation}
\Phi(t) \,\approx\, 
-  \Omega t + i \ln 
\int \hskip-2pt {dX_0 \over \sqrt{2 \pi \sigma^2}} \;
{\mathrm{e}}^{X_0^2 \over 2 \sigma^2 }\;
{\mathrm e}^{i t \,\sqrt{[\Omega \sin \theta + X_0]^2 + (\Omega \cos \theta)^2}} \, .
\label{SPA-oneqb}
\end{equation}
A convenient approximation is obtained by expanding $\Omega(X_0)$
to second order in $X_0$, which leads to~\cite{PRL05}
\begin{equation}
\label{ref:quadratic}
- i \, \Phi(t) =  i \Omega t  - {1 \over 2} \,
{(\cos \theta \, \sigma t)^2 \over 1 +  i 
\sin \theta^2 \sigma^2 t/\Omega } - {1 \over 2} \,
\ln \Big(1 +  i 
\sin \theta^2 {\sigma^2 t \over \Omega} \Big) \, .
\end{equation}
The short-times decay of coherent oscillations qualitatively depends on the
working point.
In fact, the suppression of the signal, $\exp[\MyIm \Phi(t)]$,  turns from a 
$\, \exp (- {1 \over 2} (\cos \theta \sigma t)^2)$ behavior
at $\theta \approx 0$ to a power law,  
$[1 + (\sin \theta^2 \sigma^2 t/\Omega)^2 ]^{-1/4}$, 
at $\theta \approx \pi/2$. 
In these limits Eq.(\ref{ref:quadratic}) 
reproduces known results for Gaussian $1/f$ environments. 
In particular, at $\theta = 0$ we obtain the short-times limit,
$t \ll 1/\gamma_M$, of
the exact result of Ref.\cite{kn:palma96}.   
At $\theta = \pi/2$ the short and intermediate times result
of Ref.~\cite{kn:makhlin-04} is reproduced. 
The fact that results of a diagrammatic approach 
with a quantum environment, 
as those of Ref.~\cite{kn:makhlin-04}, can be reproduced and 
generalized already at the simple SPA level makes 
the semi-classical approach quite promising. It shows that, at 
least for not too long times (but surely longer than times of interest
for quantum state processing), the quantum nature of the 
environment may not be relevant for the class of problems 
which can be treated in the Born-Oppenheimer approximation. 
Notice also that the SPA itself has surely a wide validity 
since it does not require information about the {\em dynamics} 
of the noise sources, provided they are  slow~\cite{kn:nota}.

\subsubsection{Beyond SPA: first correction}

\begin{figure}
\centering
\resizebox{1.0\columnwidth}{!}{
\includegraphics{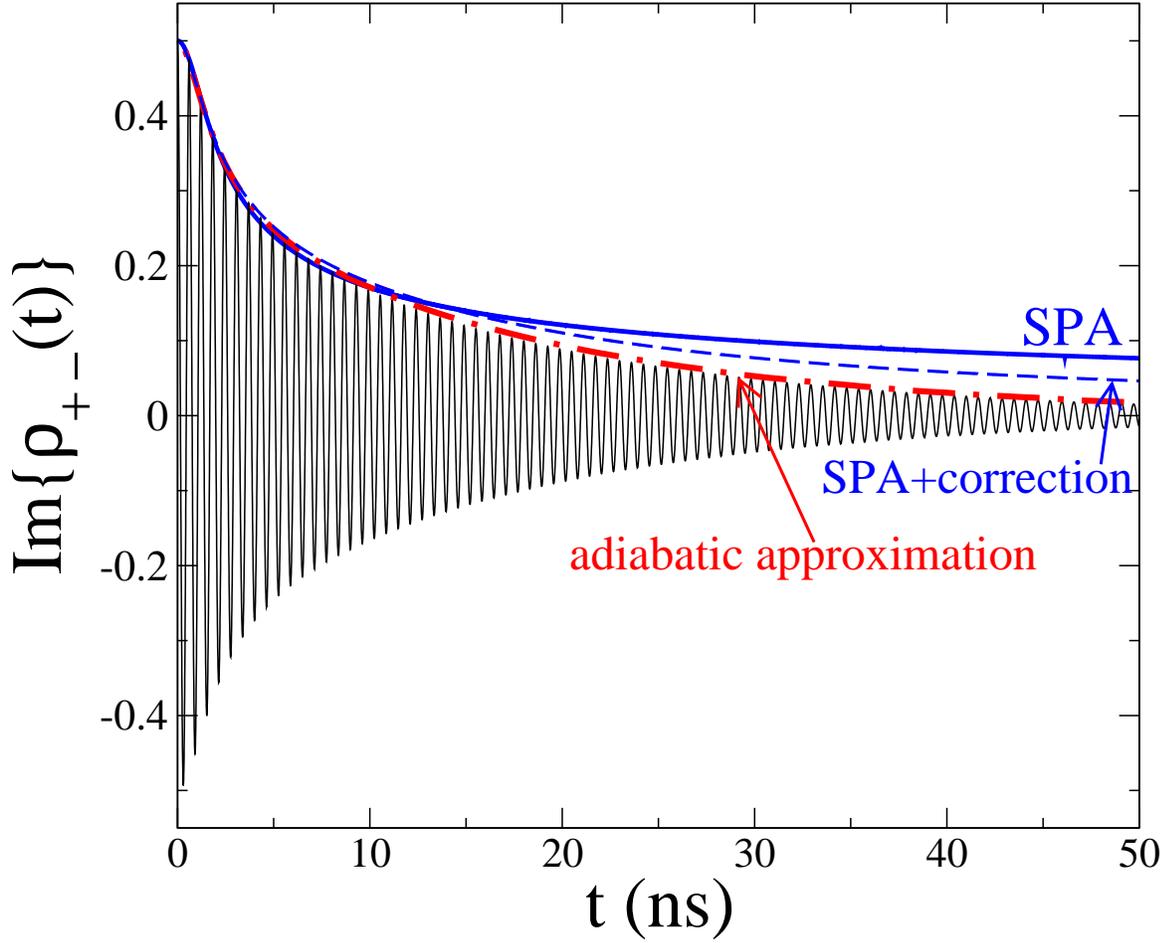}
} 
\caption{Imaginary part of the coherence $\rho_{+-}(t)$ at $\theta=\pi/2$:
Numerical simulations of an adiabatic $1/f$ environment, $S^{1/f}(\omega) \approx
 [(\pi/4) \,N_{bf} \overline{v^2} / 
\ln (\gamma_M/\gamma_m)]\,\,\omega^{-1}$, correspond to the thin black line.
Noise is produced by $n_d=250$ bistable fluctuators per decade, 
with $1/t_m=10^5$ rad/s $\le \gamma_i \le \gamma_M = 
10^9$ rad/s $< \Omega = 10^{10}$ rad/s. The coupling
$\bar{v} = 0.02 \,\Omega$ 
is appropriate to charge  devices, and corresponds to 
$S(\omega)= 16 \pi  A E_C^2/\omega$ with $A \sim 10^{-6}$~\cite{kn:zorin}. 
The  adiabatic approximation  fully
accounts for dephasing (numerical evaluation of Eq.(\ref{eq:blur}), red dot-dashed line). 
The Static Path Approximation 
(SPA) Eq.(\ref{ref:quadratic}) (blue solid line) and 
the first correction (blue dashed line) 
account for the initial suppression, and it is valid
also for times $t \gg 1/\gamma_M$. 
}
\label{figure1}
\end{figure}

Going beyond the SPA amounts to sample more accurately the stochastic process $\vec X(s)$ 
in (\ref{sampling}). The first correction to the SPA is obtained by parametrizing
the random process as follows $\vec X(s) = \vec X_0 + \frac{\vec X_t - \vec X_0}{t}s$. 
Inserting this expression in (\ref{eq:pathint-adiabatic}) and approximating $p[\vec X(s)]$
in (\ref{sampling}) with the conditional probability $p[\vec X_t,t; \vec X_0,0]$, we obtain
\begin{eqnarray}
\rho(t) 
&=&
\int  \, d \vec X_t \, d \vec X_0  \, p_2(\vec X_t,t; \vec X_0,0) \, 
\rho \big(t| \vec X_0 + \frac{\vec X_t - \vec X_0}{t}s \big)  \, .
\end{eqnarray}
The joint probability depends on the statistics of the noise sources. 
Therefore the first correction to the SPA distinguishes discrete and 
Gaussian processes.
Considering again a single qubit, the average phase (\ref{eq:blur})
at the working point $\theta=\pi/2$ for generic Gaussian noise becomes
$$
i  \Phi(t) = i \Omega t + {1 \over 2} 
\ln \Big[1 +  i 
{\sigma^2 [1 - \pi(t)] t \over \Omega} \Big]
+ {1 \over 2} 
\ln \Big[1 +  i 
{\sigma^2 \pi(t)\,t \over 3 \, \Omega} \Big] \, ,
$$
where 
$
\pi(t) = {1 \over 2 \sigma^2} \int_0^\infty \, 
(d \omega/\pi) \, S(\omega) (1- \mathrm{e}^{-i \omega t})$
is a transition probability, depending on the stochastic process.
For Ornstein-Uhlenbeck processes it reduces to the result
of Ref.~\cite{kn:averin04}. 
The first correction suggests that the SPA, in principle valid for 
$t <  1/\gamma_M$, may have  a broader validity. This is illustrated in
 Fig.(\ref{figure1}) where the adiabatic approximation (numerical evaluation of
 Eq.(\ref{eq:blur})) is compared with the exact (numerical) dynamics in the
presence of $1/f$ noise and with the analytic forms resulting from the SPA
and its first correction. 
For $1/f$ noise due to a set of bistable impurities the SPA valid also for 
$t \gg 1/\gamma_M$,    
if $\gamma_M \lesssim \Omega$. Of course 
the adiabatic  approximation is tenable if $t < T_1^A = 2/S_X^A(\Omega)$.
 
By sampling more accurately the adiabatic process $\vec X(t)$
it is possible to selectively include the statistical information needed
for the specific measurement process. For instance, echo protocols are
able to partly re-focus the signal, in other words
de-focusing described by the SPA is almost canceled by the echo
pulse sequence. The decay of the echo signal is  due to the
un-canceled dynamics of the low-frequency fluctuations and to quantum noise. 
The leading effect of adiabatic noise can be estimated by a proper  parametrization of $\vec X(t)$,
similar to the one considered in the present paragraph~\cite{preparation}.

\section{Two-qubit universal gate: multi-stage approach}

In the present Section we apply the multi-stage approach to a universal 
two-qubit gate based on a fixed coupling scheme. Capacitive and
inductive fixed couplings~\cite{coupled-th-fixed} have been used to 
demonstrate two-qubit logic gates in different JJ implementations~\cite{coupled-exp-fix}.
Entanglement is generated by tuning  single-qubit energy  spacing
to achieve mutual resonance.
To this end, during the 
gate operation at least one qubit has to be moved away from the  working point
of minimal sensitivity to parameters variations, the "optimal  point".
This has so far represented the  main drawback of the fixed-coupling scheme for 
Josephson implementations, 
with the exception of phase qubits~\cite{coupled-phase}.
More recent proposals have attempted to solve this problem  by introducing tunable coupling
schemes~\cite{coupled-th-tunable}. 
Most of them rely on additional circuit elements and gain their tunability from 
ac-driving~\cite{coupled-th-ac} or from 
"adiabatic" couplers~\cite{coupled-th-adiabatic}. Some of these
coupling schemes have been tested in experiments and 
are potentially scalable~\cite{coupled-exp}.
None of these implementations is however totally immune from imperfections.
In general, introducing additional on-chip circuits elements opens new ports to noise.
The  possibility to employ "minimal" fixed coupling schemes has been recently
reconsidered in Ref.~\cite{0906.3115}, pointing out the possibility to single out "optimal coupling"
conditions which ensure reasonable protection from $1/f$ fluctuations.

\begin{figure}
\centering
\resizebox{0.9\columnwidth}{!}{
\includegraphics{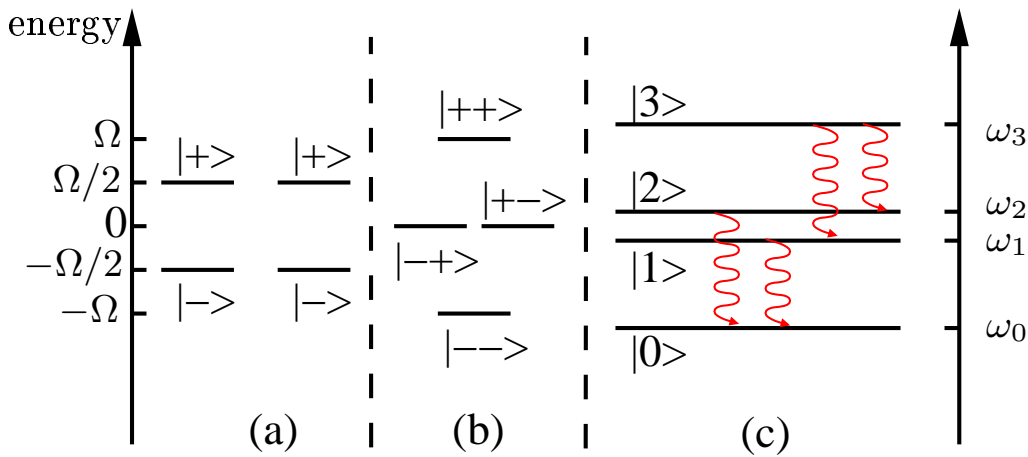}
}
\caption{(a) Level structure of the uncoupled resonant qubits;
(b) levels in the two-qubit Hilbert space and  logic basis of product states. 
(c) When the coupling is turned on $|+-\rangle$ and $|-+\rangle$ mix and 
an energy splitting $\omega_c \ll \Omega$ develops between the 
eigenstates $\{\vert 2 \rangle,\vert 1 \rangle\}$, spanning the 
SWAP subspace. Product states $|--\rangle$ and $|++\rangle$
weakly mix and split, with $2\Omega \sqrt{1 + g^2/4}$. The
eigenstates $\{\vert 0 \rangle,\vert 3 \rangle\}$ span the Z subspace.
Transverse noise (\ref{eq:Hnoise}) only mix eigenstates belonging to the
two subspaces. Dissipative transitions between subspaces are indicated by
wavy lines. At typical temperatures, $k_B T \ll \Omega$,
thermal excitation processes between subspaces can be neglected.  
}
\label{splittings}
\end{figure}

In order to implement a $\sqrt{\rm{i-SWAP}}$ gate in a fixed coupling scheme we need
two resonant qubits with a transverse coupling, as described by the
Hamiltonian 
\begin{equation}
\mathcal{H}_0 = 
 -\frac{\Omega}{2} \, \sigma_x^{(1)} \otimes \mathbb{I}^{(2)}
 -\frac{\Omega}{2}  \, \mathbb{I}^{(1)} \otimes \sigma_x^{(2)}
+ \frac{\omega_c}{2} \, \sigma_z^{(1)} \otimes \sigma_z^{(2)} 
\label{eq:hamiltonian}
\end{equation}
where $\omega_c$ is the coupling strength, and
$\sigma_x^{(i)}$ the pseudo-spin operators whose eigenstates
$ \vert \pm \rangle$ (eigenvalues $\pm1$) are the computational states 
of qubit $i$.
Eigenvalues and eigenvectors of $\mathcal{H}_0$  are given by
\vskip-6pt 
\begin{eqnarray}
\omega_0 &=& - \Omega \sqrt{1 + g^2/4} \quad , \quad 
|0 \rangle \, = \, - \sin \frac{\varphi}{2}   \,| ++ \rangle  
\, + \, \cos \frac{\varphi}{2}  \, | -- \rangle
\\
\omega_1 &=& - \omega_c/2   \quad \quad  \;, \quad \quad  \quad
|1 \rangle \, = \,\frac{1}{\sqrt{2}}\, (- |+- \rangle  \, + \,| -+ \rangle ) 
\\
\omega_2 &=& \omega_c/2  \quad \quad  \quad \; , \quad \quad  \quad \;
|2 \rangle \,= \, \frac{1}{\sqrt{2}} \,( | +- \rangle \, + \, |-+ \rangle ) \\
\omega_3 &=& \Omega \sqrt{1 + g^2/4} 
 \quad  \;, \quad  \quad
|3 \rangle \, = \,\cos \frac{\varphi}{2} \,  | ++ \rangle 
\, + \, \sin \frac{\varphi}{2}\, | -- \rangle 
\end{eqnarray}
where  $\sin \varphi = g/(2\sqrt{1+g^2/4})$,
$\cos \varphi = -1/\sqrt{1+g^2/4}$ with $g= \omega_c/\Omega$ and
we have used the shorthand notation 
$\vert \mu \nu \rangle = \vert \mu \rangle_1 \otimes
\vert \nu \rangle_2 \; \mu,\nu \in \{ +,- \}$.
The level structure of the coupled system is schematically illustrated in Fig.(\ref{splittings}).
As a result of the diagonal block structure of the Hamiltonian 
(\ref{eq:hamiltonian})
in the computational space,
the two-qubit Hilbert space is factorized in two subspaces spanned by pairs 
of eigenvectors. 
A system prepared in $| +- \rangle$, freely evolving for a time $t_E = \pi/2 \omega_c$, 
yields the entangled state $(| +- \rangle - i | -+ \rangle)/\sqrt{2}$,
corresponding to a $\sqrt{\rm{i-SWAP}}$ operation. 
The pair of states $| 1 \rangle$ and $| 2 \rangle$ 
span the subspace where the $\sqrt{\rm{i-SWAP}}$ gate is realized, which we name SWAP-subspace. 
The subspace spanned by the  pair of states 
$|0 \rangle$ and $| 3 \rangle$ is termed Z-subspace. 
Eq.(\ref{eq:hamiltonian}) models, for instance, capacitive coupled charge qubits or 
inductively coupled flux qubits~\cite{coupled-exp-fix}. 
Typically, these system can be driven by pulses 
acting along the transverse direction, $\sigma_z^{(i)}$. The same port introduces noise into
the system~\footnote{In principle noise can also couple longitudinally, i. e. via  
$\sigma_x^{(i)}$~\cite{0906.3115}.
This is the case of the charge-phase two-port design of Ref.~\cite{kn:vion}.}. 
 Here we consider 
noise sources acting transversely with respect to each qubit, i. e.  the interaction Hamiltonian takes
the form
\begin{equation}
\mathcal{H}_\mathrm{I} = -\frac{1}{2} \,  \sigma_z^{(1)} \otimes \mathbb{I}_2 \,  \hat{X}_{1}
-\frac{1}{2} \, \mathbb{I}_1 \otimes \sigma_z^{(2)} \,  \hat{X}_{2}.
\label{eq:Hnoise}
\end{equation}
Each noise component $\hat X_i$ has broadband spectrum 
$
S_{X_i}(\omega) = \frac{A}{\omega}$, 
$\omega \in \{\gamma_m, \gamma_M\} $
followed by a white flank at frequencies $\omega \geq \gamma_M$.
Correlated noise sources acting on both qubits have been addressed in \cite{NJP-special}.

If the system is initialized in the SWAP subspace, for instance in the
state $|+ \,-\rangle$, bitwise readout  
gives the qubit 1 switching probability $P^{(1)}(t)$, 
i.e. the probability that it will pass to the state $\vert - \rangle$, 
and the probability $P^{(2)}(t)$ of finding the qubit 2 
in the initial state $\vert - \rangle$. 
Cyclic anti-correlation of the probabilities signals the formation 
of the entangled state during the $\sqrt{\rm{i-SWAP}}$ operation.
In terms of the two qubit reduced density matrix in the eigenstate basis 
the switching probabilities  read
\begin{eqnarray}
\!\!\!\!\!\!\!\!\!\!\!\!\!\!\!\!\!\!\!\!\!\!\!\!\!\!\!P^{(1)}(t) 
 =\, \langle - | {\rm Tr}_2 \rho(t) | - \rangle
 &=& \frac{1}{2} \left [ \, \rho_{11}(t) + \rho_{22}(t) \right ]
+ \rho_{00}(t)
+\left [ \, \rho_{33}(t) - \rho_{00}(t) \right ] \sin^2 \frac{\varphi}{2} 
\nonumber \\
&& + {\rm Re}[\rho_{12}(t)] +  {\rm Re} [\rho_{03}(t)] \sin \varphi
\label{eq:psw1} \\
\!\!\!\!\!\!\!\!\!\!\!\!\!\!\!\!\!\!\!\!\!\!\!\!\!\!\!P^{(2)}(t) 
 = \langle - | {\rm Tr}_1 \rho(t) | - \rangle & =&
\frac{1}{2} \left [ \, \rho_{11}(t) + \rho_{22}(t) \right ] 
+  \rho_{00}(t)  
+  \left [ \, \rho_{33}(t) - \rho_{00}(t) \right ] \sin^2 \frac{\varphi}{2} 
\nonumber \\ 
&&-  {\rm Re}[\rho_{12}(t)] + {\rm Re}[\rho_{03}(t)] \sin \varphi.
\label{eq:psw2}
\end{eqnarray}
The matrix elements entering the above probabilities can be evaluated in
the multi-stage approach.

\subsection{Multi-stage approach}

We split the interaction as in Eq.(\ref{int-plit}), 
$\hat X_i = \hat X_i^Q +  X_i(t)$.
Low-frequency noise is treated in  the adiabatic and longitudinal approximation.
In addition we limit the analysis to the SPA. 
We denote with $\omega_i(X_1,X_2)$ 
the eigenvalues of $\mathcal{H}_0 + \mathcal{H}_\mathrm{I} $ with 
$\hat X_i \to \{X_{i}\}$.

\begin{figure}
\centering
\resizebox{1.0\columnwidth}{!}{
\includegraphics{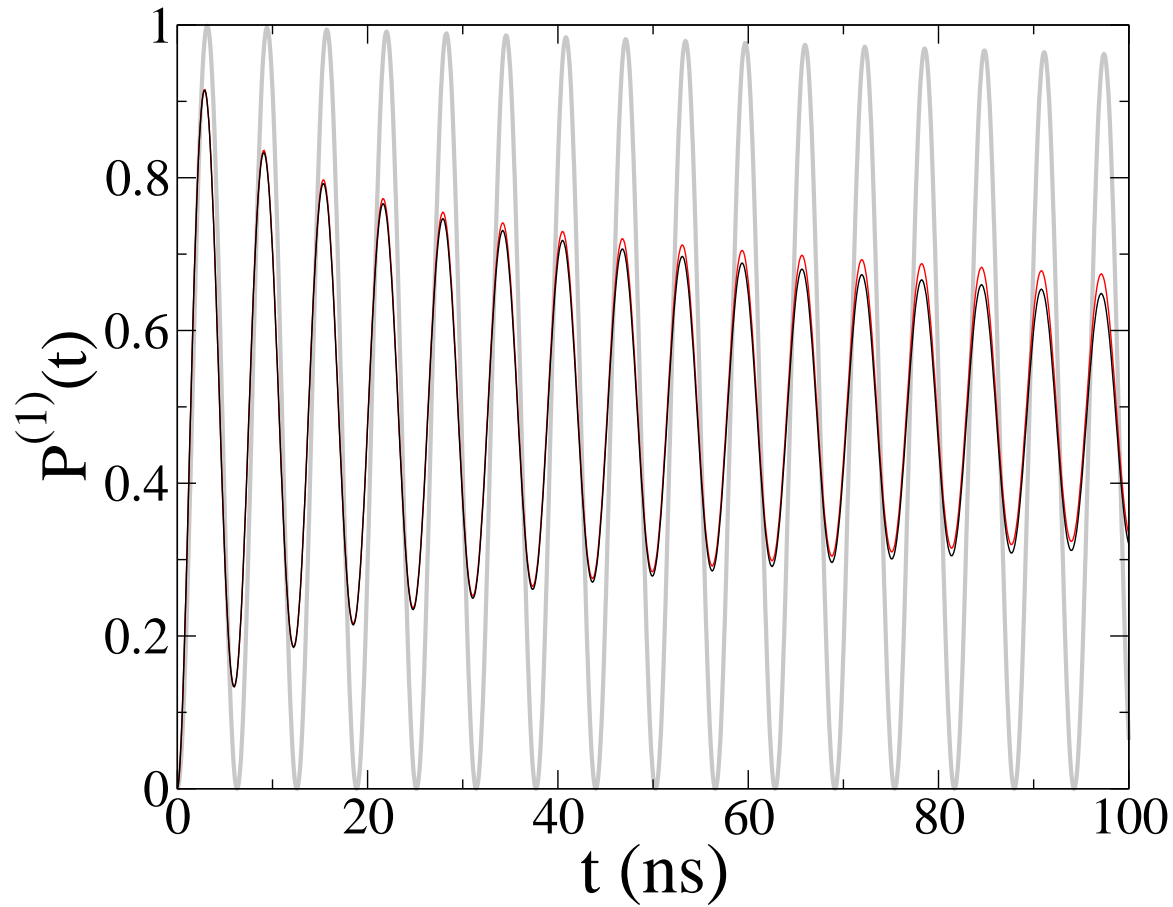}
} 
\caption{Interplay of adiabatic and quantum noise in the switching probability 
of qubit 1. Adiabatic noise in the SPA is parametrized by  $\sigma/\Omega = 0.08$.
The  effect of low-frequency noise  is shown in red,
the effect of white noise, $S_{x_i}(\omega) \approx 8 \times 10^5$s$^{-1}$, in gray.
Interplay of adiabatic and quantum noise is shown by the black line.
The interaction strength is $\omega_c/\Omega= 0.01$. 
}
\label{figure2}
\end{figure}

{\em First stage: elimination of quantum noise}
Quantum noise is traced out by solving the Born-Markov master equation
for the reduced density matrix~\cite{PhysicaE} with eigenvalues parametrically dependent on the
random fields $\{X_{i}\}$. 
In the system eigenstate basis~\footnote{We are here disregarding effects due to the
instantaneous eigenstates of $\mathcal{H}_0 + \mathcal{H}_\mathrm{I} $ with 
$\hat X_i \to \{X_{i}\}$} 
and performing the secular approximation (to be self-consistently checked) it takes the
standard form~\cite{kn:cohen}:
\begin{eqnarray}
\dot \rho_{ii}(t) &=& - \sum_{m \neq i} \Gamma_{im} \, \rho_{ii}(t) + 
\sum_{m \neq i} \Gamma_{mi} \, \rho_{mm}(t) 
\label{eq:populations}
\\
\dot \rho_{ij}(t) &=& - (i  \omega_{ij} + \widetilde\Gamma_{ij} ) 
\,\rho_{ij}(t) \,, 
\label{eq:coherences}
\end{eqnarray}
where $ \omega_{ij}= \omega_i(X_1,X_2) -\omega_j(X_1,X_2)$.
The rates $\Gamma_{lm}$, $\tilde\Gamma_{ij}$  depend  on the real
parts of the lesser and greater Green's functions,
describing  emission (absorption) rates to (from)  the quantum
reservoirs~\cite{book}.  In addition, for white quantum noise energy shifts are vanishing and
do not appear in Eq. (\ref{eq:coherences}).
Because of the symmetry of (\ref{eq:hamiltonian}), dissipative transitions inside each
-~SWAP or Z~- subspace are forbidden. The allowed inelastic energy exchange processes 
are evidenced in Fig.(\ref{splittings}). The only independent emission rates are  
$\Gamma_{10}=\Gamma_{32}$, $\Gamma_{20}=\Gamma_{31}$. 
They read
\begin{eqnarray}
\begin{array}{ll}
\Gamma_{10} = \frac{1}{8} \, (1+\sin \varphi) \, [ C_{X_1}(\omega_{10}) 
+  C_{X_2}(\omega_{10})]
\\
\Gamma_{20} = \frac{1}{8} \, (1-\sin \varphi) \, [ C_{X_1}(\omega_{20}) 
+  C_{X_2}(\omega_{20})] \, ,
\end{array}
\label{eq:ratesgeneral}
\end{eqnarray}
where the absorption rates,
$C_{X_i}(\omega) = \frac{2 \, S_{X_i}^Q(\omega)}{1+ \exp{(- \omega/ k_B T)}}$,
are related to the spectrum of quantum noise, $S_{X_i}^Q(\omega)$.
Emission rates have the same form with $C_{X_i}(\omega_{lm})$
replaced by  $ C_{X_i}(-\omega_{lm})$. 
For the considered initial condition, $|+-\rangle$, 
the only non vanishing elements
of the reduced density matrix in the eigenbasis of ${\mathcal H}_0$ are the populations
and the SWAP coherence, $\rho_{21}(t)$. 
For independent quantum noise sources  acting on the two qubits, the SWAP decay rate reads
$\widetilde \Gamma_{12} = \frac{1}{2} \, [\Gamma_{10} + \Gamma_{01} + \Gamma_{20} + \Gamma_{02} ]
= \frac{1}{2} \, [\Gamma_1 + \Gamma_2] $, 
where $\Gamma_i = \Gamma_{i0} + \Gamma_{0i}$, $i=1,2$, are relaxation rates of the SWAP levels.
From Eq.(\ref{eq:coherences}) we get
\begin{equation}
\rho_{21}(t) = - \frac{1}{2} \, e^{i \omega_{21}(X_1,X_2) t} \, e^{-\widetilde \Gamma_{12}t } \,.
\end{equation}
The secular approximation is valid provided that 
$\omega_{21} \approx  \omega_c \gg \widetilde \Gamma_{12}$. This condition is fulfilled,
for instance, for white noise levels extrapolated from single charge-phase qubit
experiments, as it can be evinced from the slow decay due to quantum noise
reported in  Fig.(\ref{figure2}), gray line.

Equations (\ref{eq:populations}) for the populations do not
decouple even in the secular limit. General solutions are quite
lengthy. Here we report the approximate forms in the small temperature
limit with respect to the uncoupled qubits splittings, 
$k_B T \ll \Omega$. In this case level $3$ remains unpopulated and
\begin{eqnarray}
\!\!\!\!\! \!\! \!\!\!\!\! \!\!\!\!\!\rho_{00}(t) &\approx& 1 - \frac{1}{2} [e^{- \Gamma_{20} t} + e^{-\Gamma_{10} t}] 
\quad , \quad 
\rho_{11}(t)  \approx  \frac{1}{2} \, e^{- \Gamma_{10} t} \quad , \quad
\rho_{22}(t) \approx \frac{1}{2} \, e^{- \Gamma_{20} t} \, .
\label{populations}
\end{eqnarray}

\begin{figure}
\centering
\resizebox{1.0\columnwidth}{!}{
\includegraphics{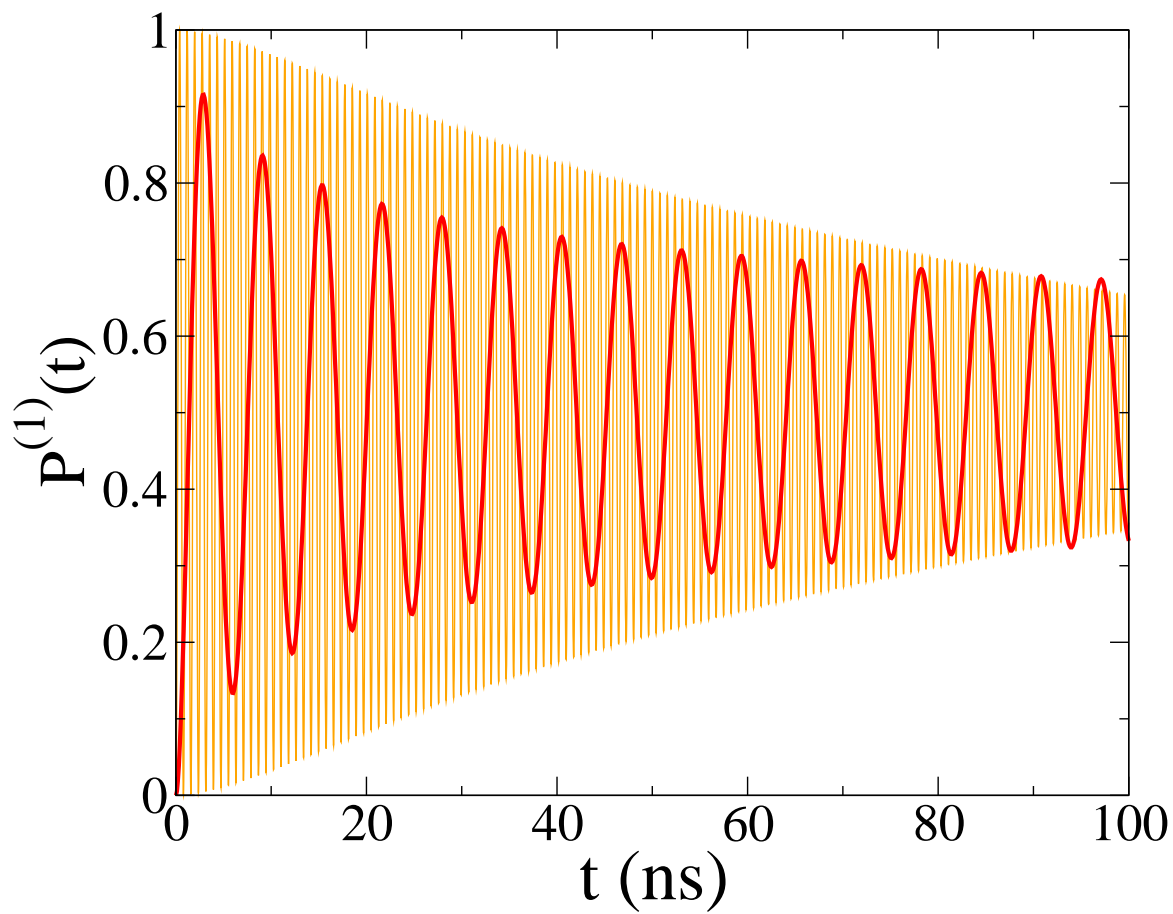}
} 
\caption{Effect of adiabatic noise under optimal tuning: switching probability 
of qubit 1 in the presence of low-frequency noise parametrized by $\sigma/\Omega = 0.08$
for  "optimal coupling" $\omega_c/\Omega= 0.08$ (orange) and non-optimal
tuning $\omega_c/\Omega= 0.01$ (red). 
}
\label{figure3}
\end{figure}

{\em Second stage: elimination of adiabatic noise}
We now consider the effect of low-frequencies in the
adiabatic, longitudinal and Static Path approximations. Populations are
unaffected by adiabatic noise, whereas the SWAP coherence, $\rho_{12}(t)$,
has to be averaged  over $X_{i}$. Here we disregard the  
negligible  $X_{i}$ dependence of the rates $\Gamma_{ij}$
via $\omega_{ij}$.
Under these simplifying assumption,  we are left with the following average  
\begin{equation}
\rho_{21}(t) \approx -\frac{1}{2} \,  e^{-\widetilde \Gamma_{12}t } \, 
\int d X_1 d X_2  \, p(X_1)p(X_2)  \,e^{ i  \omega_{21}(X_1,X_2) t}.
\label{eq:staticint}
\end{equation}
with 
$p(X_i) = \frac{1}{\sqrt{2 \pi}\sigma_i} \exp{[- \frac{X_i^2}{2\sigma_i^2}]}$.
The SWAP splitting can be estimated by treating in perturbation theory 
$\mathcal{H}_\mathrm{I} $, with $\hat X_i \to \{X_{i}\}$, with 
respect to $\mathcal{H}_0$. This leads to
\begin{eqnarray}
\omega_{21}(X_1,X_2) &\approx& \omega_c -\frac{\omega_c}{2 \Omega^2}  (X_1^2+X_2^2) 
+  \frac{\omega_c}{8 \Omega^4}  (1+  \frac{\omega_c^2}{\Omega^2})
(X_1^4+ 6 X_1^2 X_2^2 + X_2^4) 
\nonumber \\
&+& \frac{1}{8\omega_c \Omega^2}(X_1^2-X_2^2)^2 . 
\end{eqnarray}
The average in Eq. (\ref{eq:staticint}) can be evaluated in analytic form and gives~\cite{0906.3115}
\begin{equation}
\rho_{12}(t) = - \frac{1}{2}  \,  e^{-\widetilde \Gamma_{12}t } \, 
\frac{ \Omega }{2 \sigma^2} \sqrt{\frac{2 i \omega_c}{\pi  t}} \, 
e^{i \omega_c t+ h(t)} \, K_0[ h(t)] 
\label{eq:SPA-SWAP}
\end{equation}
where 
$h(t)=  i \omega_c/t \, (\Omega^2/\sigma^2 +i \omega_c t)^2/(4 \Omega^2)$,
and $K_0[h]$  is the K-Bessel function of order zero~\cite{abramowitz}. We
considered equal standard deviations for noise sources acting on
both qubits, $\sigma_i \equiv \sigma$. 
Inserting (\ref{populations}) and (\ref{eq:SPA-SWAP})  
in (\ref{eq:psw1}) and (\ref{eq:psw2}) we obtain the
switching probabilities in the multi-stage approach.  
Out of phase oscillations
signals two-qubit states anti-correlations 
and follows from
$P^{({}^{1}_{2})}(t) = P(t) \pm {\rm Re}[\rho_{12}(t)]$, with 
$P(t)= - \frac{1}{2} \, \cos \varphi \,
\left [ \rho_{11}(t) + \rho_{22}(t)  \right ] + 
\cos^2\left (\frac{\varphi}{2}\right ) \approx  - \frac{1}{4} \, \cos \varphi \,
\left [ e^{- \Gamma_{10}t} + e^{- \Gamma_{20}t} \right ] + 
g^2/4 $.

The efficiency of the gate results from the interplay of quantum and 
adiabatic noise. High-frequency noise levels expected from single qubit 
experiments~\cite{ithier} weakly affect the switching probability, 
whose decay is mainly due to low-frequency noise, Fig. (\ref{figure2}).
Remarkably, considerable recovery of short-times oscillation amplitude may
be achieved by an optimal choice of the coupling strength, 
$\omega_c \approx \sigma$~\cite{0906.3115}. This regime is illustrated in
Fig.(\ref{figure3}).

\section{Conclusions}

In this article we presented a road-map to treat broadband noise 
typical of solid state nanodevices. The introduced multi-stage approach 
allows to obtain reasonable approximations by systematically 
including only the relevant information on the complex environment, 
out of the huge parametrization which would be required for
a microscopic description. 
Since the environment is in general long-time correlated, 
the required information depends on the specific protocol.

The predictions obtained with the present approach  are in agreement with observations  
in various single qubit JJ implementations and in  
different protocols~\cite{kn:falci-varenna,PRL05}.
We extended the procedure to deal with complex solid-state architectures.
This is a required step in order to predict efficiency and possibly
appropriately design of nanodevices for  quantum information processing.
Both because of the complexity of architectures and of
the unavoidable broadband nature of solid state noise,
theoretical tools allowing systematic and controlled
approximations are particularly valuable.

 As an illustrative case, we performed a simplified analysis
of the effects and interplay of low and high frequency noise components
in a two-qubit gate in a fixed coupling scheme.
Our results points out that efficient operations in the solid state require
an accurate preliminary characterization of the noise spectral characteristics and
tuning appropriately the device working point.

\ack
We acknowledge support from the EU-EuroSQIP (IST-3-015708-IP).


\section*{References}
\begin{thebibliography}{99}
\bibitem{Nielsen} 
M. Nielsen, I. Chuang, 
		"Quantum Computation and Quantum Information", 
		Cambridge Univ. Press, 2005.
\bibitem{single-semic} T. Hyashi {\em et al.}, Phys. Rev. Lett. {\bf 91}, 
		226804 (2003); J. R. Petta {\em et al.}, Phys. Rev. Lett. 
		{\bf 93}, 186802 (2004);
		J.R. Petta {\em et al.}, Science {\bf 309}, 2180 (2005); 
		J. Gorman, D.G. Hasko, and D.A. Williams, Phys. Rev. Lett. 
		{\bf 95}, 090502  (2005);
		F. H. L. Koppens {\em et al.}, Nature {\bf 442}, 766 (2006).
\bibitem{single-super}
	Y. Nakamura 
	{\em et al.}, {Nature} {\bf 398}, {786} (1999);
	Y. Yu {\em et al.},  
	{Science} {\bf 296}, {889} (2002);
	J.M. Martinis {\em et al.},  
	{Phys. Rev. Lett.} {\bf 89}, {117901} (2002);
	I. Chiorescu {\em et al.},  
	Science, {\bf 299}, 1869, (2003);
        T. Yamamoto {\em et al.},
       	Nature {\bf 425}, 941 (2003);
        S. Saito {\em et al.}, Phys. Rev. Lett. {\bf 93}, 037001 (2004);
	J. Johansson {\em et al.}, ibid. {\bf 96}, 127006 (2006);
	F. Deppe {\em et al.}, Nat. Phys. {\bf 4}, 686 (2008);
\bibitem{kn:vion} 
	D. Vion {\em et al.}, 
	{Science} {\bf 296}, {886} (2002).
\bibitem{trasmon} J. Koch et al. Phys. Rev. A {\bf 76}, 042319 (2007); 
	J. A. Schreier {\em et al.}, Phys. Rev. B {\bf 77}, 180502(R) (2008).	
\bibitem{circuitQED} D. I. Schuster, {\em et al.}
	Nature  {\bf 431}, 162  (2004); 
	D. I. Schuster {\em et al.} 
	Nature {\bf 445}, 515  (2007);
	M. G\"oppl {\em et al.} 
	Nature {\bf 454}, 315  (2008); 
	Lev S. Bishop {\em et al.} 
	Nature Physics {\bf 5}, 105  (2008).
\bibitem{kn:nakamura-echo} Y. Nakamura  {\em et. al}, 
	{ Phys. Rev. Lett.} {\bf 88}, {047901} (2002).
\bibitem{kn:falci-varenna} G. Falci, E. Paladino, R. Fazio,
	in {\em Quantum Phenomena of Mesoscopic Systems}, B. L. Altshuler and
	V. Tognetti Eds.,  Proc. of the International School ``Enrico Fermi'', Varenna 2002, 
       IOS Press (2003), cond-mat/0312550.
\bibitem{PRL05} G. Falci {\em et al.}, 	Phys. Rev. Lett. {\bf 94}, 167002 (2005).
\bibitem{slitcher} C. P. Slichter  \textit{Principles of Magnetic Resonance}, Springer-Verlag, Berlin
		(1996)
\bibitem{delsing08} S. Kafanof {\em et al.} Phys. Rev. B, {\bf 78}, 125411 (2008). 
\bibitem{nak-spectrum} O. Astafiev 	{\em et al.},
		Phys. Rev. Lett. {\bf 93}, 267007 (2004). 
\bibitem{ithier} G. Ithier  {\em et al.},  Phys. Rev. B
		{\bf 72}, 134519 (2005).
\bibitem{martinis04} R.W. Simmonds {\em et al.},~Phys. Rev. Lett. {\bf 93}, 077003 (2004); 
	 	K.B.~Cooper {\em et al.}, ibid. {\bf 93}, 180401 (2004).
\bibitem{mooji} J. Eroms {\em et al.}, Appl. Phys. Lett. {\bf 89}, 122516 (2006). 
\bibitem{Delft}  C. H. Van der Wal, 
 	{\em et al.}, Science {290}, 773777 (2000).
\bibitem{kn:duty} T. Duty {\em et al.},  Phys. Rev. B {\bf 69}, 140503(R) (2004); 
\bibitem{Altshuler} Y. M. Galperin {\em et al.}
	Phys. Rev. Lett. {\bf 96}, 097009 (2006);
	J Bergli, Y M Galperin and B L Altshuler, New J. Phys. {\bf 11},  025002 (2009).
\bibitem{poletto} F. Chiarello private communication 2008. Experiment based on the setup of 
		S.Poletto {\em et al.} New J. Phys. {\bf 11}, 013009 (2008).
\bibitem{siewert} R. Fazio, G. M. Palma, and J. Siewert
	Phys. Rev. Lett. {\bf 83}, 5385 (1999). 
\bibitem{weissman} {M.B.\ Weissman}, {Rev.\ Mod.\ Phys.} {\bf 60}, {537} (1988). 
\bibitem{coherent} J. Claudon {\em et al.},  
	 	Phys. Rev. B {\bf 76}, 024508 (2007);
	L. Tian and R.W. Simmonds, Phys. Rev. Lett. {\bf 99}, 137002 (2007);
	F. Deppe {\em et al.} Phys. Rev. B {\bf 76}, 214503 (2007);
	Z. Kim,  {\em et al.} Phys. Rev. B {\bf 78}, 144506 (2008);
	A. Lupascu {\em et al.}
	arXiv:0810.0590.
\bibitem{zagoskin} A. M. Zagoskin {\em et al.} 
	Phys. Rev. Lett. {\bf 97}, 077001 (2006).
\bibitem{neeley08} M.  Neeley, {\em et al.}
Nature Physics {\bf 4}, 523 (2008).
\bibitem{PRB08} E. Paladino  {\em et al.} 
	Phys. Rev. B  {\bf 77}, 041303(R)(2008).
\bibitem{coherent-th} S. Ashab {\em et al.}, New J. Phys. {\bf 8}, 103 (2006);
	 N. P Oxtoby {\em et al.}
	New J. Phys. {\bf 11}, 063028  (2009).
\bibitem{kn:cohen} C. Cohen-Tannoudji, J. Dupont-Roc and G. Grynberg
		{\em Atom-Photon Interactions}, Wiley-Interscience (1993).
\bibitem{PRL02} E. Paladino	{\em et al.},
	Phys. Rev. Lett. {\bf 88}, 228304 (2002).
\bibitem{NJP-special} Correlated noise and cross-talk effects have been addressed
in A. D'Arrigo {\em et al.} NJP {\bf 10}, 115006 (2008).
\bibitem{preparation} G. Falci {\em et al.} in preparation.
\bibitem{kn:zurek} W. Zurek, Physics Today {\bf 44}, 36 (1991).
\bibitem{kn:palma96} G. M. Palma, K.A. Suominen, A. K. Ekert, 
		Proc. R. Soc. London A, {\bf 452}, {567} (1996). 
\bibitem{kn:cory-et-al-1998} D. G. Cory {\em et al.} Phys. Rev. Lett. {\bf 81}, 2152 (1998).
\bibitem{kn:isham-95} C. Isham, {\em Lectures on Quantum Theory: Mathematical and Structural Foundations},
		Imperial College Press, London (1995).
\bibitem{kn:makhlin-04}  Y. Makhlin, A. Shnirman, Phys. Rev. Lett. {\bf 92}, 178301 (2004).
\bibitem{kn:zorin} A.B. Zorin {\em et al.}, {Phys.\ Rev.\ B} {\bf 53}, {13682} (1996). 
\bibitem{kn:nota} Eq. (\ref{ref:quadratic}) is also valid for Ornstein-Uhlenbeck processes, see
		\cite{kn:averin04}. For random telegraph noise the discrete nature of
		the process modifies this result see E. Paladino {\em et al.}
		Adv. Sol. State Phys., {\bf 43}, 747 (2003)
\bibitem{kn:averin04} K. Rabenstein, V.A. Sverdlov, D.V. Averin
		JETP Letters {\bf 79}, 646 (2004).
\bibitem{coupled-th-fixed} {Yu.} Makhlin {\em et al.},
	{ Nature} {\bf 398}, 305 (1999);
	J.~Q. You {\em et al.},
	{ Phys. Rev. Lett.} {\bf 89}, 197902 (2002).	
\bibitem{coupled-exp-fix} 
	{Yu}.~A. Pashkin {\em et al.},{ Nature} {\bf 421}, 823 (2003);
	A.~J. Berkley {\em et al.},  {Science} {\bf 300}, 1548 (2003);
	T.~Yamamoto {\em et al.}, { Nature} {\bf 425}, 941 (2003);
	R.~McDermott {\em et al.},  { Science} {\bf 307}, 1299 (2005);
	J.~B. Majer {\em et al.},
	{Phys. Rev. Lett.} {\bf 94}, 090501 (2005);
	M. Steffen {\em et al.}, {Science} {\bf 313}, 1423 (2006);
	J.H. Plantenberg {\em et al.},
	{ Nature} {\bf 447}, 836 (2007).
\bibitem{coupled-phase} M. A. Sillanp\"a\"a, J. I. Park and  R. W. Simmonds
	Nature {\bf 449}, 438 (2007).
\bibitem{coupled-th-tunable} 
	  D.V. Averin, C. Bruder,
	{Phys. Rev. Lett.} {\bf 91}, 057003 (2003);
	A. Blais  {\em et al.}, 
	{ibid.} {\bf 90}, 127901 (2003);
	F. Plastina, G. Falci, {Phys. Rev. B} {\bf 67}, 224514 (2003);
	B.Plourde {\em et al.},
	 {ibid.} {\bf 70}, 140501(R) (2004);
	A.O. Niskanen, Y. Nakamura, J.S. Tsai,
       { ibid.} {\bf 73}, 094506 (2006);
       P. Bertet, C.~J.  Harmans, J.E. Mooij,
	{ ibid.} {\bf 73}, 064512 (2006);
	Y-D Wang, A. Kemp, K. Semba, {ibid.} 
	{\bf 79}, 024502 (2009).
\bibitem{coupled-th-ac}  
	C.~Rigetti, A.~Blais, and M.~Devoret, {Phys. Rev. Lett.} {\bf 94}, 240502 (2005);
	{Yu-xi} Liu {\em et al.}, 
	{ibid.} {\bf 96}, 067003 (2006);
	G. S. Paraoanu, Phys. Rev. B 74, 140504(R) (2006).
\bibitem{coupled-th-adiabatic} 
	T.~V. Filippov {\em et al.}, { IEEE Trans. Appl. Supercond.} {\bf 13}, 1005
	(2003);
	J.~Lantz {\em et al.}, { Phys. Rev. B} {\bf 70}, 140507(R) (2004);
	A.~{Maassen van den Brink} {\em et al.}, 
	{ New J. Phys.} {\bf 7}, 230 (2005).
\bibitem{coupled-exp}
	 A. Izmalkov {\em et al.}, 
	 {Phys. Rev. Lett.} {\bf 93}, 037003 (2004);
	M. Grajcar   {\em et al.},  PRB {\bf 72}, 02503 (2005);
	T. Hime {\em et al.}, 
	 { Science} {\bf 314},  1427 (2006);
	A. O. Niskanen {\em et al.}, 
	{ Science}  {\bf 316} 723 (2007); 
	 A. O. Niskanen {\em et al.}, {Nature} {\bf 447}, 386 (2007);
	J. Majer {\em et al.}, {ibid.} {\bf 449}, 443 (2007);
	S. H. W. van der Ploeg  {\em et al.}, Phys. Rev. Lett. {\bf 98}, 057004 (2007);
	A. Fay {\em et al.},
	ibid. {\bf 100}, 187003 (2008);
	A. O. Niskanen {\em et al.},
	 Phys. Rev. B {\bf 77}, 064505 (2008);
	K. Harrabi {\em et al.},
	Phys. Rev. B 79, 020507 (2009).
\bibitem{0906.3115} E. Paladino, A. Mastellone, A. D'Arrigo, G. Falci, arXiv:0906.3115 
\bibitem{book} U. Weiss, "Quantum dissipative systems", Third edition, World Scientific 2008.
\bibitem{abramowitz} M. Abramowitz, I. A. Stegun ``Handbook of Mathematical
Functions'', Dover (1965).
\bibitem{PhysicaE} E. Paladino {\em at al.}  Physica E in press (2009).
\end{thebibliography}
\end{document}